\documentclass[twocolumn,showpacs,prl]{revtex4}

\usepackage{graphicx}
\usepackage{dcolumn}
\usepackage{bm}

\begin{document}

\setlength{\baselineskip}{0.4cm} \addtolength{\topmargin}{1.5cm}

\title{Experimental study of the effect of disorder on subcritical crack growth dynamics}

\author{O. Ramos$^{1}$}
\author{P.-P. Cortet$^{2}$}
\author{S. Ciliberto$^{3}$}
\author{L. Vanel$^{1}$}

\affiliation{ $^1$Institut Lumi\`{e}re Mati\`{e}re, UMR5306 Universit\'{e}  Lyon 1-CNRS, Universit\'{e} de Lyon, 69622 Villeurbanne, France\\
$^2$Laboratoire FAST, CNRS, Univ. Paris Sud, UPMC Univ. Paris 06, 91405 Orsay cedex, France\\
$^3$Universit\'{e} de Lyon, Laboratoire de Physique,
CNRS UMR5672, Ecole Normale Sup\'{e}rieure de Lyon, 69364 Lyon Cedex 07, France}

\date{\today}

\begin{abstract}

The growth dynamics of a single crack in a heterogeneous material under subcritical loading is an
intermittent process; and many features of this dynamics have been shown
to agree with simple models of thermally activated rupture. In
order to better understand the role of material heterogeneities in this
process, we study the subcritical propagation of a crack in a sheet of paper in the presence of a
distribution of small defects such as holes. The experimental data obtained for two different distributions of holes are discussed in the light of models that
predict the slowing down of crack growth when the disorder in the
material is increased; however, in contradiction with these theoretical predictions, the experiments result in longer lasting cracks in a more ordered scenario. We argue that this effect is \emph{specific} to subcritical crack dynamics and that the weakest zones between holes at close distance to each other are responsible both for the acceleration of the crack dynamics and the slightly different roughness of the crack path.
\end{abstract}


\maketitle

{\it Introduction.} When a brittle material is loaded beyond a critical stress it breaks almost instantaneously;  however, even a subcritical loading is generally able to fracture the system, but in a time-dependent manner. Subcritical fracture in disordered materials usually follows a rather complex dynamics, with the occurrence of large --power law
like-- distributions of rupture event sizes, observed both in
experiments and numerical models \cite{Petri et al 1994, Garcimartin
et al 1997, Santucci et al 2004, Maloy et al 2006, Koivisto et al 2007, Kloster et al 2006, Minozzi et al 2003}. Experimental evidences of such distributions are based either on indirect
observations, like the detection of acoustic emissions associated to
rupture events \cite{Petri et al 1994,Garcimartin et al 1997,Koivisto et al 2007}, or on
direct observations of the propagation of a crack front
\cite{Santucci et al 2004, Maloy et al 2006}.
Power law statistics is obtained in models describing the rupture
of a disordered material through a threshold dynamics where the time scale of rupture is either set by the loading rate or by an imposed creep/damage law \cite{Kloster et al 2006, Minozzi et al 2003, Zapperi et al 2005, Kun et al 2003, Pradhan et al 2010,Halasz et al 2012}. Experimentally, the rupture process has been proposed to be triggered by thermal noise (for a review see \cite{Vanel et al 2009}). This is actually a very important practical case for structures that are submitted to an essentially
constant load such as buildings, which are constantly evolving towards thinner and lighter designs. Following this idea, a number of
theoretical works have been developed recently so as to describe the
rupture dynamics driven by thermal noise in materials with a
distribution of local rupture thresholds \cite{Scorretti et al 2001, Saichev and Sornette 2005, Guarino et al 2006,
Cortet et al 2006, Kierfeld and Vinokur 2006}. It has been found that, for a given mean rupture threshold, the amplitude of disorder (defined as the standard deviation of the rupture thresholds) can diversely affect the rupture dynamics. It can
accelerate the dynamics when it involves many spatially diffuse
rupture events \cite{Scorretti et al 2001,
Saichev and Sornette 2005, Guarino et al 2006} or on the
contrary slow it down when it involves the growth of a
main dominant crack \cite{Cortet et al 2006, Kierfeld and Vinokur
2006}. Controlling experimentally the
disorder in a material is a rather difficult task which limits
tremendously the possibility to test these theoretical predictions. For example, Dalmas \emph{et al.} \cite{Dalmas et al 2008} have analyzed the effects of the scale of the disorder in the fracture process of three-dimensional samples, but their studies are limited to the roughness of the crack surfaces. In
this Letter, we propose a method to control material disorder by weakening a paper sample along the path of a main crack through the addition of small holes. The dynamics and structure of the crack is analyzed under two different spatial distributions of holes, while keeping constant the damage introduced to the material, i.e., the averaged density of holes.

\begin{figure*}[t!]
\vspace{-0.3cm}
\includegraphics[width=7in, height=2.40in]{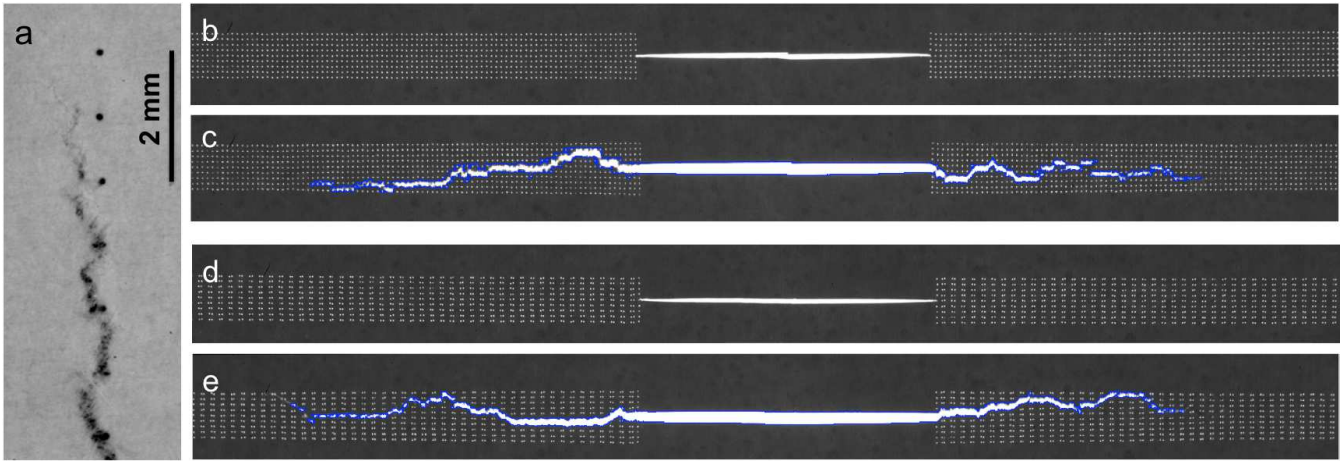}
\vspace{-0.3cm} \caption{\label{cracks} (color online) (a) Passage of a crack across a line of 1mm--equispaced holes. (b-e) Views of
the initial and final crack (a few moments before the
sample breaks apart). (b, c) in the case of a $unimodal$ spacing of
0.5 mm between holes. (d, e) in the case of a $bimodal$ hole
spacing, alternating between $0.25~$mm and $0.75~$mm in each line.}
\vspace{-0.3cm}
\end{figure*}

{\it Experimental procedure.} We use fax paper samples from Alrey having
a thickness of $50~\mu$m. An initial crack of 3 cm length is formed at the
center of each sample of dimensions $21~$cm$~\times~24~$cm. Holes are
pierced through the sample using an acupuncture needle of diameter
$120~\mu$m actuated by a three-axial computer-controlled platform. The first samples we studied had a series of holes regularly spaced by a distance $d=1~$mm and aligned along the direction of the initial crack (Fig.~\ref{cracks} a). However, the crack deviations due to the intrinsic disorder of the sample were often driving the crack away from the holes. In order to guarantee that the crack will interact with holes along its whole path, we reduced the hole spacing to $d=0.5~$mm and added four clones of this line parallel to the original alignment on each side of it, keeping a spacing of $0.5~$mm between them (Figs.~\ref{cracks} b, c).  The samples with an $ordered$ pattern of holes will be further labeled as the $unimodal$ configuration.

In order to build a disordered configuration, a simple way could have been to randomly change the position of the holes along each line of the $unimodal$ configuration. In this process, in order to avoid the collapse of two holes into a single defect, the amplitude of these random displacements should be limited, for instance, between -0.125~mm and 0.125~mm. As a result, the inter-hole distances would be randomly distributed between 0.25~mm and 0.75~mm. However, in this disordered configuration many inter-hole distances would still be close to 0.5~mm, like in the unimodal case. To make a clear difference with the unimodal configuration, we have chosen to use only the two extreme values of inter-hole distances and build a line of holes with non-random alternating distances of 0.25~mm and 0.75~mm creating a 1~mm periodicity. Eight cloned lines are added to the samples, similarly to the ordered configuration (Figs.~\ref{cracks} d, e). This {\it less ordered} configuration will be labeled as the $bimodal$ configuration. We emphasize that theoretically a bimodal distribution is a disorder enough to create a global slowing down of the crack dynamics as can be inferred from \cite{Cortet et al 2006}. The disorder created in the samples will actually be a disorder on the stress field rather than on the rupture thresholds. However, the effect on the dynamics will be very similar since thermally activated rupture is controlled by the difference between the local rupture threshold and the local stress \cite{Vanel et al 2009}. Due to stress concentration around the holes, we expect that, in the bimodal case, the dynamics will be faster between holes closer to each other and slower between holes farther apart.

Experiments are performed by applying a constant force perpendicularly to the initial crack direction. The temperature is controlled and kept constant at 35 $\pm$ 0.1$\,^{\circ}{\rm C}$ and the relative humidity is kept below 5$\%$ during all experiments. Crack growth is
followed using a high resolution camera (ImperX 11 Megapixels) at 5 frames/s.
Crack contours are extracted using a digital image analysis routine
that is precise enough to separate the crack from the background of
holes. Figure~\ref{cracks}(c, e) shows in blue examples of extracted crack contours
superimposed with the raw image for both configuration of holes.

\begin{figure}[b!]
\vspace{-0.4cm}
\includegraphics[ width=3.2in]{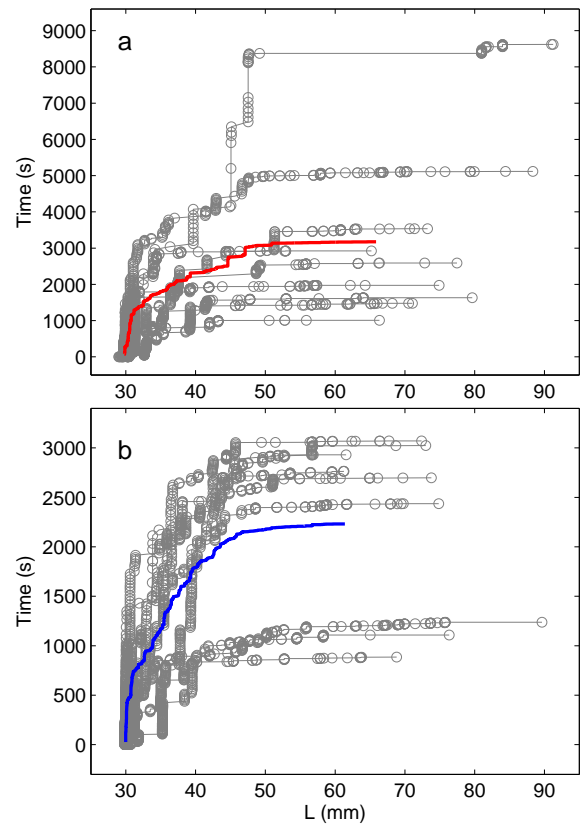}
\vspace{-0.4cm}
 \caption{\label{data} (color online) Time vs. crack length (considering only  the contribution parallel to the main crack direction). (a) Unimodal samples. (b)
Bimodal samples. The solid lines correspond to the ensemble averaged curves.}
\end{figure}

\begin{figure}[t!]
\includegraphics[ width=3.2in]{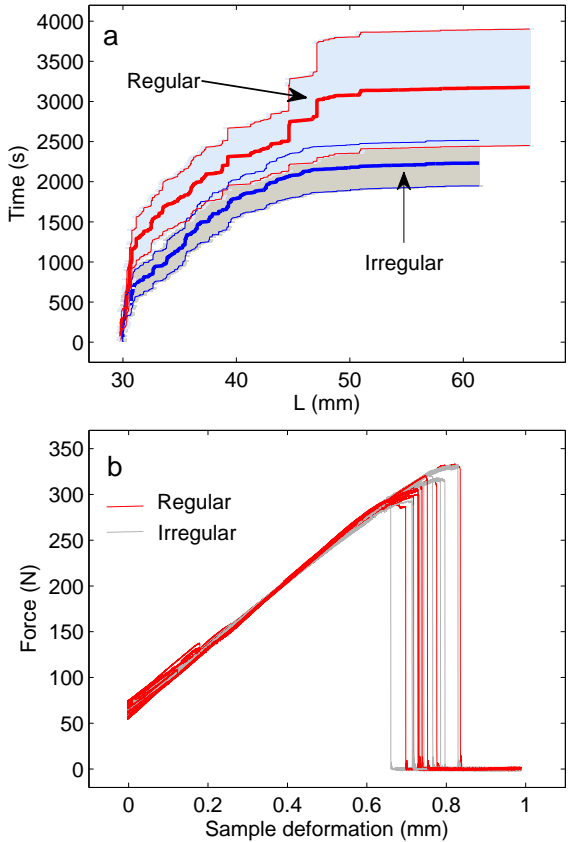}
\vspace{-0.2cm}
 \caption{\label{averages} (color online) (a) Ensemble averages of the time vs. crack length from Fig.~\ref{data}. The shadowed areas correspond to the standard deviation of the data. They are limited by a thin solid line. (b) Force vs. elongation for eighteen samples (nine unimodals and nine bimodals) submitted to a loading rate of 45 $\mu$m/s.}
 \vspace{-0.5cm}
\end{figure}

{\it Results.} In Fig.~\ref{data}, we plot the extracted crack growth curves for eighteen
independent experiments performed in the same conditions with either
the unimodal (a) or the bimodal (b) configuration. Each curve shows that the crack advances by a series of sudden jumps occuring after the crack has been pinned for a certain time, until it reaches a critical length separating the subcritical behavior from an abrupt rupture. The critical lengths, estimated as in \cite{Santucci et
al 2004}, are comparable for both configurations: $\langle L_{c}\rangle=76.4\pm 9.0~$mm for the unimodal case and $\langle L_{c}\rangle= 72.4 \pm 8.5~$mm for the bimodal one. However, the rupture times, at which the critical lengths have been attained, are more dispersed and present (in ensemble average) higher values in the unimodal configuration ($\langle t_{max}\rangle=3209~$s) than in the bimodal one ($\langle t_{max}\rangle= 2238~$s), as shown in Fig.~\ref{averages}a. The fact that subcritical cracks propagate (in average) slower in the more ordered scenario is the main result of this work and the rest of this Letter will discuss the corresponding spatio-temporal features.

First, in order to verify that this result is specific to the subcritical behavior, eighteen samples, similar to those used in Fig.~\ref{data}, were broken by imposing a deformation ramp with a velocity of 45~$\mu$m/s. We measure \emph{no noticeable departure between the two configurations} (Fig.~\ref{averages}b), with critical rupture thresholds equal to $\langle F_{max}\rangle = 312 \pm 14~$N and $\langle F_{max}\rangle=$ 310 $\pm$ 13~N for the single and bimodal spacing respectively. We do not observe here a dependence of the critical force on disorder as predicted in \cite{Alava et al 2008}.

\begin{figure}[b!]
\vspace{-0.4cm}
\includegraphics[width=3.2in]{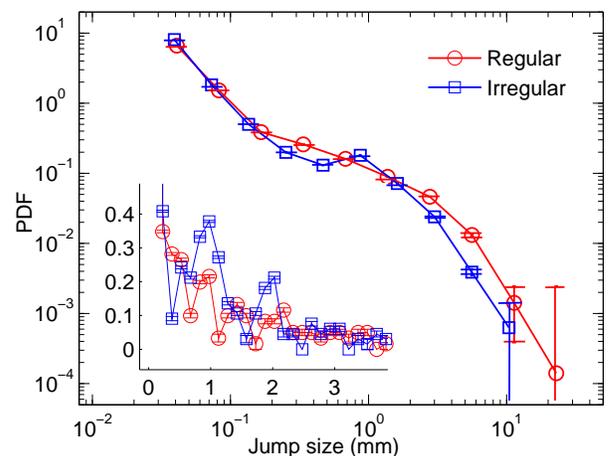}
\vspace{-0.4cm}
 \caption{\label{jumps} Probability distribution function (PDF) of jump sizes for the two
configurations. A characteristic jump size appears in the bimodal case and is better observed in the inset showing the PDF in lin-lin scale. Error bars are estimated as $N^{-1/2}$ times the value of the PDF, where $N$ is the number of samples.}
\end{figure}

If one analyzes the statistics of crack jumps, taking into account
all the data in Fig.~\ref{data}, one finds a rather broad distribution similar to the one reported earlier for samples with no holes \cite{Santucci et
al 2004}. However, one noticeable feature distinguishes the unimodal and bimodal configurations (Fig.~\ref{jumps}), namely the appearance of a peak
at a jump size close to $1~$mm for the bimodal hole pattern. This peak is a robust feature when varying the box width used to build the distribution and thus reveals a preferentially selected jump distance. Interestingly, this size
corresponds to the periodicity of the bimodal pattern that is
double the one of the unimodal pattern. As seen in the inset of
Fig.~\ref{jumps}, the next clear peak in jump sizes for the bimodal
pattern occurs for $2~$mm, i.e. twice the basic
pattern periodicity. In the case of a unimodal pattern, it
seems that there is also a selection of jump with sizes multiples of the periodicity,
here $0.5~$mm. The inset of Fig.~\ref{jumps} shows indeed small peaks close to jump values twice, three times and four times the basic pattern period. However, it does not show clearly a peak for jump sizes of $0.5~$mm but only an inflexion point in the distribution.

An explanation of the 1~mm jumps in the bimodal case could be that once two holes distant by 0.75~mm are connected, it is much
easier to connect the following hole at a distance $0.25~$mm, so as to
create effectively a jump size of $1~$mm. Making a na\"{\i}ve calculation, if we consider as $zero$ the time spent by the crack in crossing all the 0.25~mm interhole regions and assume the crack dynamics is the same in the $0.5~$mm and $0.75~$mm interhole regions, the crack in the bimodal pattern should be 25$\%$ faster than the one in the unimodal pattern ($3209~$s$~\times~(1-0.25) = 2407~$s). This simple analysis provides a very reasonable estimate of the experimental rupture time measured in the bimodal configuration. This reasoning is corroborated by the fact no crack tips can be detected in the images in the 0.25~mm interhole regions, which confirms a fracture velocity much larger in these regions.

\begin{figure}[t!]
\vspace{-0.4cm}
\includegraphics[width=3.2in]{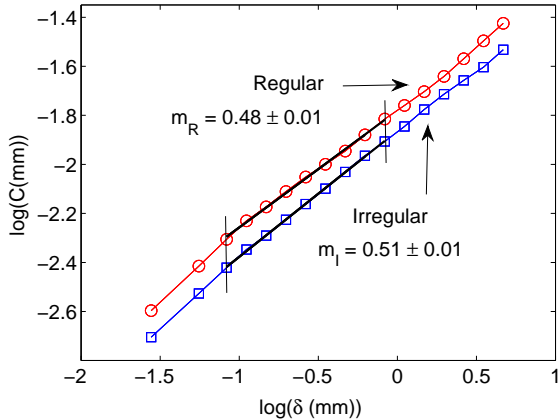}
\vspace{-0.2cm}
 \caption{\label{roughness} Logarithm of structure functions of the crack path as a function of
scale, fitted by lines in the regime corresponding to the sub-interhole distances. In order to separate the curves, 0.1 units has been added to the upper one. }
\vspace{-0.6cm}
\end{figure}

The previous discussion indicates that, in the bimodal configuration, the crack dynamics between holes at close distance to each other ($0.25~$mm) is significantly faster than between holes further apart ($0.75$~mm), which could have an impact on the roughness of the crack path \cite{Mallick et al 2007}. We have analyzed the roughness of the path followed by the crack in both sample configurations. For a crack path defined by the function $y(x)$, where $y$ is the crack deviation from a straight growth direction along $Ox$, we used the
structure function:  $C(\delta)=\langle [y(x+\delta)- y(x)]^2 \rangle _x ^{1/2}$ (Fig.~\ref{roughness}). In order to capture the crack roughness at the interhole scale, the analysis focused on a range standing  from $80~\mu$m (three pixels) to $800~\mu$m. In this range, the exponent in the unimodal configuration is: m$~=0.48\pm 0.01$ and in the bimodal one: m$~=0.51\pm 0.01$. These exponents are much smaller than the one measured in the same paper samples without holes in the subcritical rupture regime ($\approx$0.65 \cite{Mallick et al 2007}). As a crack deviating from a straight path by random walk steps would have a roughness exponent of $1/2$, the exponent measured here at the interhole distance scale suggests that the crack is exploring randomly the pattern of holes, as if attracted by them. Furthermore, although the roughness exponents are almost the same in both configurations, the slight increase observed for the bimodal configuration suggests that the crack is a little more often attracted by holes in the forward direction than for the unimodal case, which is again compatible with the assumption made above that holes separated by a distance of $0.25$~mm are very easily connected by the growing crack.


{\it Conclusion.} In this Letter, we have reported experiments on the subcritical propagation of a crack in paper sheets in which an artificial disorder was created by adding patterns of holes along the crack path. Two types of samples with different spatial distributions of holes, but the same hole density, were used. Rupture dynamics turns out to be slower in the case of a unimodal pattern of holes than in the case of the bimodal one. This effect is \emph{specific} to subcritical crack dynamics since the critical rupture thresholds of unimodal and bimodal samples are the same. It is however contrary to recent theoretical models predicting a slower rupture dynamics for a single crack as disorder increases  due to the leading role played by strong heterogeneities located just ahead of the advancing crack \cite{Cortet et al 2006,
Kierfeld and Vinokur 2006}. The overall acceleration observed in the bimodal configuration experiments can be quantified by assuming that the rupture dynamics between holes close to each other becomes suddenly faster than between holes farther apart. This dynamics between very close holes can actually be considered almost instantaneous which creates a stronger tendency for the crack to move straight ahead by connecting two close holes ($0.25~$mm) and could explain the slightly higher roughness exponent observed in the bimodal configuration. Accordingly, the quantitative analysis suggests that the crack dynamics is almost the same between holes separated by $0.75~$mm and by $0.5~$mm. It means that the disorder induced in the bimodal sample does not seem able to produce strong regions of higher resistance to rupture which in turn explains why the theoretical models fail to properly predict the crack dynamics in our experiments. The fact that the acceleration of the crack in the binomial case is specific to the subcritical behavior, also carries an important practical relevance: measuring the resistance of a disordered material by imposing a deformation ramp leading to the failure of the sample is arguably not the best way to test a structure built for supporting a constant load.

We thank J.-P. Bouchaud and S. Santucci for insightful discussions and R. Planet for technical support.

\end{document}